\documentclass[twocolumn,showpacs,preprintnumbers,amsmath,amssymb,times]{revtex4}
\usepackage{graphicx}
\usepackage{dcolumn}
\usepackage{bm}
\usepackage{amsmath}

\begin{document}
\title{Epidemic spreading on heterogeneous networks with identical infectivity}
\author{Rui Yang}
\author{Bing-Hong Wang}
\author{Jie Ren}
\author{Wen-Jie Bai}
\author{Zhi-Wen Shi}
\author{Wen-Xu Wang}
\author{Tao Zhou}
\email{zhutou@ustc.edu}

\affiliation{ Department of Modern Physics and Nonlinear Science
Center, University of Science and Technology of China, Hefei 230026,
PR China}
\date{\today}

\begin{abstract}
In this paper, we propose a modified
susceptible-infected-recovered (SIR) model, in which each node is
assigned with an identical capability of active contacts, $A$, at
each time step. In contrast to the previous studies, we find that
on scale-free networks, the density of the recovered individuals
in the present model shows a threshold behavior. We obtain the
analytical results using the mean-field theory and find that the
threshold value equals $1/A$, indicating that the threshold value
is independent of the topology of the underlying network. The
simulations agree well with the analytic results. Furthermore, we
study the time behavior of the epidemic propagation and find a
hierarchical dynamics with three plateaus. Once the highly
connected hubs are reached, the infection pervades almost the
whole network in a progressive cascade across smaller degree
classes. Then, after the previously infected hubs are recovered,
the disease can only propagate to the class of smallest degree
till the infected individuals are all recovered. The present
results could be of practical importance in the setup of dynamic
control strategies.
\end{abstract}


\maketitle
\section{Introduction}
Many real-world systems can be described by complex networks,
ranging from nature to society. Recently, power-law degree
distributions have been observed in various networks
\cite{BA,albert}. One of the original, and still primary reasons
for studying networks is to understand the mechanisms by which
diseases and other things, such as information and rumors spread
over \cite{Pastor2003,Zhou2006}. For instance, the study of
networks of sexual contact \cite{186,Liljeros2001,303} is helpful
for us to understand and perhaps control the spread of sexually
transmitted diseases. The susceptible-infected-susceptible (SIS)
\cite{Pastor2001a,Pastor2001b}, susceptible-infected-removed (SIR)
\cite{May2001,Moreno2002}, and susceptible-infected (SI)
\cite{M1,Zhou2005,Vazquez2006} models on complex networks have
been extensively studied recently. In this paper, we mainly
concentrate on the behaviors of SIR model.

The standard SIR style contains some unexpected assumptions while
being introduced to the scale-free (SF) networks directly, that
is, each node's potential infection-activity (infectivity),
measured by its possibly maximal contribution to the propagation
process within one time step, is strictly equal to its degree. As
a result, in the SF network the nodes with large degree, called
hubs, will take the greater possession of the infectivity. This
assumption cannot represent all the cases in the real world, owing
to that the hub nodes may be only able to contact limited
population at one period of time despite their wide acquaintance.
The first striking example is that, in many existing peer-to-peer
distributed systems, although their long-term communicating
connectivity shows the scale-free characteristic, all peers have
identical capabilities and responsibilities to communication at a
short term, such as the Gnutella networks \cite{p1,p2}. Second, in
the epidemic contact networks, the hub node has many
acquaintances; however, he/she could not contact all his/her
acquaintances within one time step \cite{zhutou}. Third, in some
email service systems, such as the Gmail system schemed out by
Google , their clients are assigned by limited capability to
invite others to become Google-users \cite{google}. The last, in
network marketing processes, the referral of a product to
potential consumers costs money and time(e.g. a salesman has to
make phone calls to persuade his social surrounding to buy the
product). Therefore, generally speaking, the salesman will not
make referrals to all his acquaintances \cite{market}. Similar
phenomena are common in our daily lives. Consequently, different
styles of the practical lives are thirst for research on it and
that may delight us something interesting which could offer the
direction in the real lives.

\section{The model}
First of all, we briefly review the standard SIR model. At each
time step, each node adopts one of three possible states and
during one time step, the susceptible (S) node which is connected
to the infected (I) one will be infected with a rate $\beta$.
Meanwhile, the infected nodes will be recovered (R) with a rate
$\gamma$, defining the effective spreading rate
$\lambda=\beta/\gamma$. Without losing generality, we can set
$\gamma=1$. Accordingly, one can easily obtain the probability
that a susceptible individual $x$ will be infected at time step
$t$ to be
\begin{equation}
\lambda_x(t)=1-(1-\lambda)^{\theta(x,t-1)},
\end{equation}
where $\theta(x,t-1)$ denotes the number of contacts between $x$
and its infected neighbors at time $t-1$. For small $\lambda$, one
has
\begin{equation}
\lambda_x(t)\approx\lambda\theta(x,t-1).
\end{equation}
In the standard SIR network model, each individual will contact
all its neighbors once at each time step, and therefore the
infectivity of each node is equal to its degree and $\theta(x,t)$
is equal to the number of $x$'s infected neighbors at time $t$.
However, in the present model, we assume every individual has the
same infectivity $A$. That is to say, at each time step, each
infected individual will generate $A$ contacts where $A$ is a
constant. Multiple contacts to one neighbor are allowed, and
contacts not between susceptible and infected ones, although
without any effect on the epidemic dynamics, are also counted just
like the standard SIR model. The dynamical process starts with
randomly selecting one infected node. During the first stage of
the evolution, the number of infected nodes increases. Since this
also implies a growth of the recovered population, the ineffective
contacts become more frequent. After a while, in consequence, the
infected population begins to decline. Eventually, it vanishes and
the evolution stops. Without special statement, all the following
simulation results are obtained by averaging over $100$
independent runs for each of $300$ different realizations, based
on the Barab\'asi-Albert (BA) \cite{BA} network model.

\begin{figure}
\scalebox{0.80}[0.80]{\includegraphics{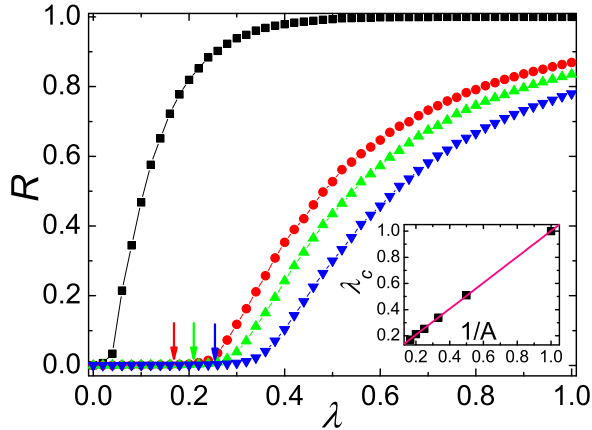}}
\caption{\label{fig:epsart} (color online) $R(\infty)$ as a
function of the effective spreading rate $\lambda$ on BA networks
with $\langle k\rangle=12$, $N=2000$. The black line represents
the case of standard SIR model, and the blue, green and red curves
represent the present model with $A=4$, 5 and 6, respectively. The
arrows point at the critical points obtained from simulations. One
can see clear from the inset that the analytic results agree well
with the simulations.}
\end{figure}

\section{Simulation and results}
Toward the standard SIR model, Moreno \emph{et al.} obtained the
analytical value of threshold $\langle k\rangle/\langle k^2
\rangle$ \cite{Moreno2002}. Similarly, we consider the time
evolution of $I_k(t)$, $S_k(t)$ and $R_k(t)$, which are the
densities of infected, susceptible, and recovered nodes of degree
$k$ at time $t$, respectively. Clearly, these variables obey the
normalization condition:
\begin{equation}
I_k(t)+S_k(t)+R_k(t)=1.
\end{equation}
Global quantities such as the epidemic prevalence are therefore
expressed by the average over the various connectivity classes;
$i.e.$, $R(t)=\sum_{k}P(k)R_k(t)$. Using the mean-field
approximation, the rate equations for the partial densities in a
network characterized by the degree distribution $P(k)$ can be
written as:
\begin{eqnarray}
&&\frac{dI_k(t)}{dt}=-I_k(t)+\lambda
k S_k(t)\sum_{k'}\frac{P(k'|k)I_{k'}(t)A}{k'}, \\
&&\frac{dS_k(t)}{dt} =-\lambda
k S_k(t)\sum_{k'}\frac{P(k'|k)I_{k'}(t)A}{k'}, \\
&&\frac{dR_k(t)}{dt}=I_k(t),
\end{eqnarray}
where the conditional probability $P(k'|k)$ denotes the degree
correlations that a vertex of degree $k$ is connected to a vertex
of degree $k'$. Considering the uncorrelated network,
$P(k'|k)=k'P(k')/\langle k\rangle$, thus Eq. (4) takes the form:
\begin{equation}
\frac{dI_k(t)}{dt}=-I_k(t)+\frac{\lambda k}{\langle
k\rangle}S_k(t)\sum_{k}AP(k)I_{k}(t).
\end{equation}
The equations (4-6), combined with the initial conditions
$R_k(t)=0$, $I_k(0)=I_k^0$, and $S_k(0)=1-I_k^0$, completely
define the SIR model on any complex network with degree
distribution $P(k)$. We will consider in particular the case of a
homogeneous initial distribution of infected nodes, $I_k^0=I^0$.
In this case, in the limit $I^0\rightarrow0$, we can substitute
$I_k(0)\simeq0$ and $S_k(0)=1$. Under this approximation and by
taking the similar converting like from Eq. (4) to Eq. (7), Eq.
(5) can be directly integrated, yielding
\begin{equation}
S_k(t)=e^{-\lambda k \phi(t)}.
\end{equation}
where the auxiliary function $\phi(t)$ is defined as:
\begin{equation}
\phi(t)=\int_0^t\frac{\sum_{k}AP(k)I_{k}(t)}{\langle
k\rangle}=\frac{\sum_{k}AP(k)R_{k}(t)}{\langle k\rangle}.
\end{equation}
Focusing on the time evolution $\phi(t)$, one has
\begin{eqnarray}
\frac{d\phi(t)}{dt}&=&\frac{\sum_{k}AP(k)I_{k}(t)}{\langle k\rangle}
\\ &=&\frac{\sum_{k}AP(k)(1-R_k(t)-S_k(t))}{\langle k\rangle} \\
&=&\frac{A}{\langle
k\rangle}-\phi(t)-\frac{\sum_{k}AP(k)S_{k}(t)}{\langle k\rangle}
\\ &=&\frac{A}{\langle k\rangle}-\phi(t)-\frac{\sum_{k}AP(k)e^{-\lambda k \phi(t)}}{\langle
k\rangle}.
\end{eqnarray}
Since $I_k(\infty)=0$ and consequently
$\lim_{t\rightarrow\infty}{d\phi(t)}/{dt}=0$, we obtain from Eq.
(13) the following self-consistent equation for $\phi_\infty$ as
\begin{equation}
\phi_\infty=\frac{A}{\langle k\rangle}(1-\sum_{k}P(k)e^{-\lambda k
\phi_\infty}).
\end{equation}
The value $\phi_\infty=0$ is always a solution. In order to have a
non-zero solution, the condition
\begin{equation}
\frac{A}{\langle k\rangle}\frac{d(1-\sum_{k}P(k)e^{-\lambda k
\phi_\infty})}{d\phi_\infty}\mid_{\phi_\infty=0}>1
\end{equation}
must be fulfilled, which leads to
\begin{equation}
\frac{A\lambda}{\langle k\rangle}\sum_{k}kP(k)=\lambda A>1.
\end{equation}
This inequality defines the epidemic threshold
\begin{equation}
\lambda_c=\frac{1}{A},
\end{equation}
below which the epidemic prevalence is null, and above which it
attains a finite value. Correspondingly, the previous works about
epidemic spreading in SF networks present us with completely new
epidemic propagation scenarios that a highly heterogeneous
structure will lead to the absence of any epidemic threshold.
While, now, it is $1/A$ instead (see the simulation and analytic
results in Fig. 1). Furthermore, we can also find that the larger
of identical infectivity $A$, the higher of density of $R(\infty)$
for the same $\lambda$ from Fig. 1.

\begin{figure}
\scalebox{0.80}[0.80]{\includegraphics{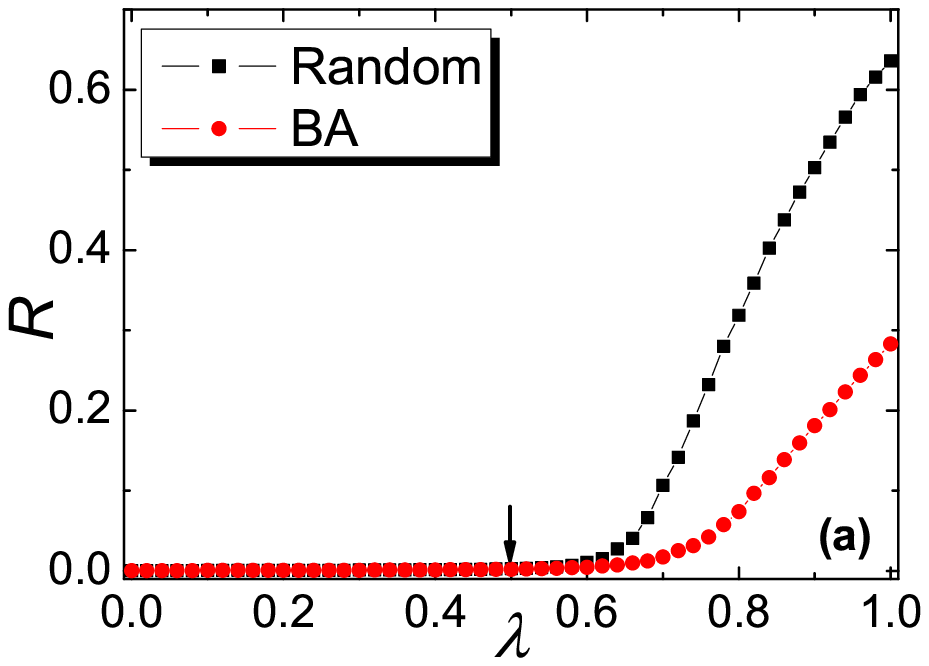}}
\scalebox{0.80}[0.80]{\includegraphics{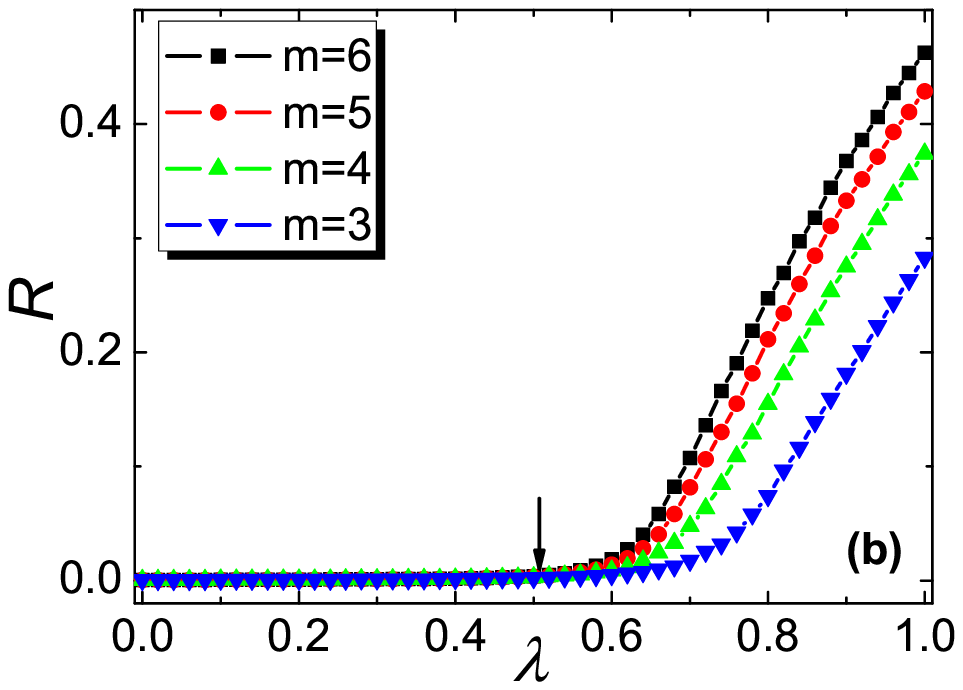}}
\scalebox{0.80}[0.80]{\includegraphics{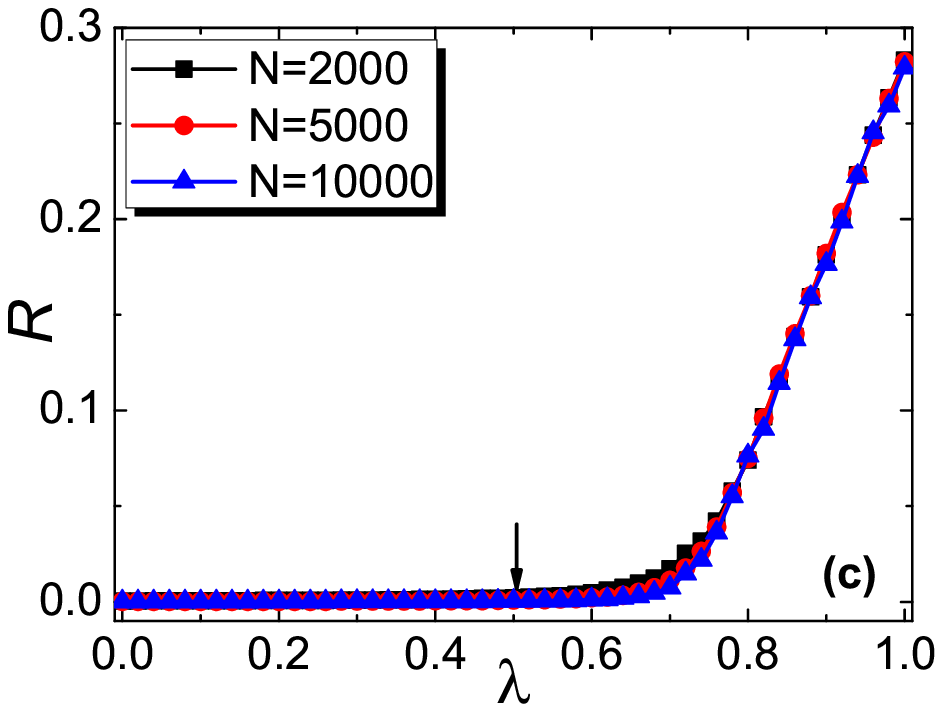}}
\caption{\label{fig:epsart}(color online) $R(\infty)$ as a
function of the effective spreading rate $\lambda$ on BA and
random networks with $\langle k\rangle=6$ (a), BA networks for
different attachment number $m$ ($m=\langle k\rangle /2$) (b), and
BA networks with different size $N$ (c). In figure (a) and (b),
the network size is $N=2000$, and in all the above three plots,
the infectivity is $A=2$. }
\end{figure}

From the analytical result, $\lambda_c=1/A$, one can see that the
threshold value is independent of the topology if the underlying
network is valid for the mean-field approximation \cite{ex}. To
further demonstrate this proposition, we next compare the
simulation results on different networks. From Fig. 2, one can
find that the threshold values of random networks, BA networks
with different average degrees, and BA networks with different
sizes are the same, which strongly support the above analysis.
Note that, in the standard SIR model, there exists obviously
finite-size effect \cite{May2001,Pastor2002}, while in the present
model, there is no observed finite-size effect (see Fig. 3(c)).

\section{Velocity and hierarchical spread}
For further understanding the spreading dynamics of the present
model, we study the time behavior of the epidemic propagation.
Originated from the Eq. (8), $S_k(t)=e^{-\lambda k \phi(t)}$,
which result is valid for any value of the degree of $k$ and the
function $\phi(t)$ is positive and monotonically increasing. This
last fact implies that $S_k$ is decreasing monotonically towards
zero as time goes on. For any two values $k>k'$, and whatever the
initial conditions $S_k^0$ and $S_{k'}^{0}$ are, there exists a
time $t'$ after which $S_k(t)<S_{k'}{(t)}$. A more precise
characterization of the epidemic diffusion through the network can
be achieved by studying some convenient quantities in numerical
spreading experiments in BA networks. First, we measure the
average degree of newly recovered nodes at time $t$, which is
equal to the average degree of newly infected nodes at time $t-1$,

\begin{eqnarray}
\langle k_{R}{(t)}
\rangle&=&\frac{\sum_{k}{k[R_k(t)-R_k(t-1)]}}{R(t)-R(t-1)} \\
&=&\frac{\sum{_k}{k I_k(t-1)}}{I(t-1)} \\ &=&\langle k_{inf}{(t-1)}
\rangle.
\end{eqnarray}

\begin{figure}
\scalebox{0.80}[0.80]{\includegraphics{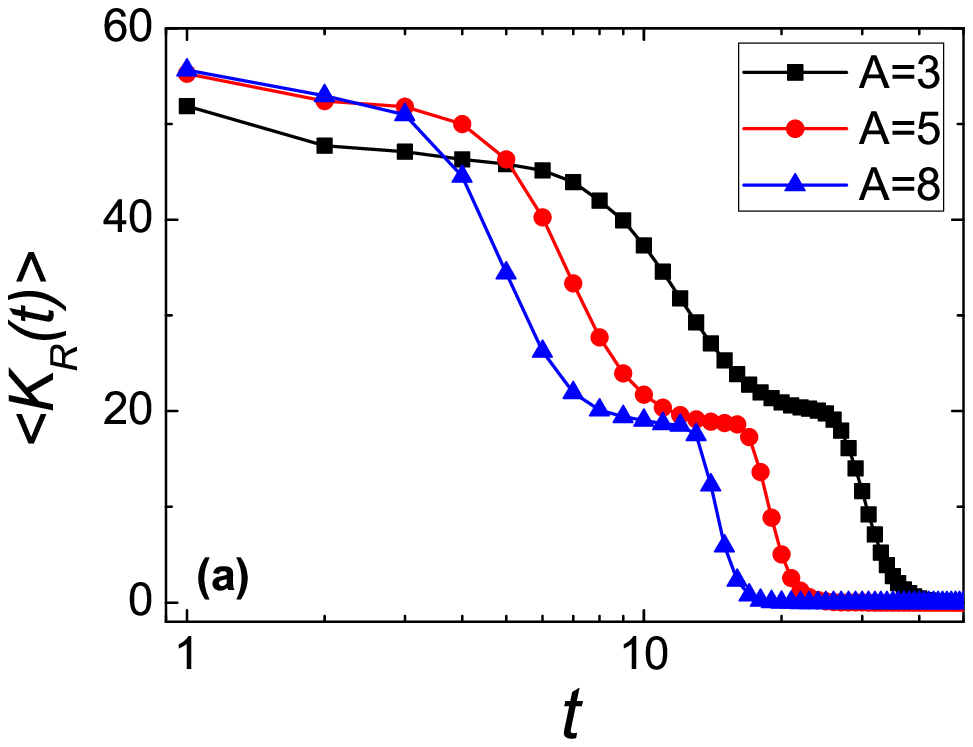}}
\scalebox{0.80}[0.80]{\includegraphics{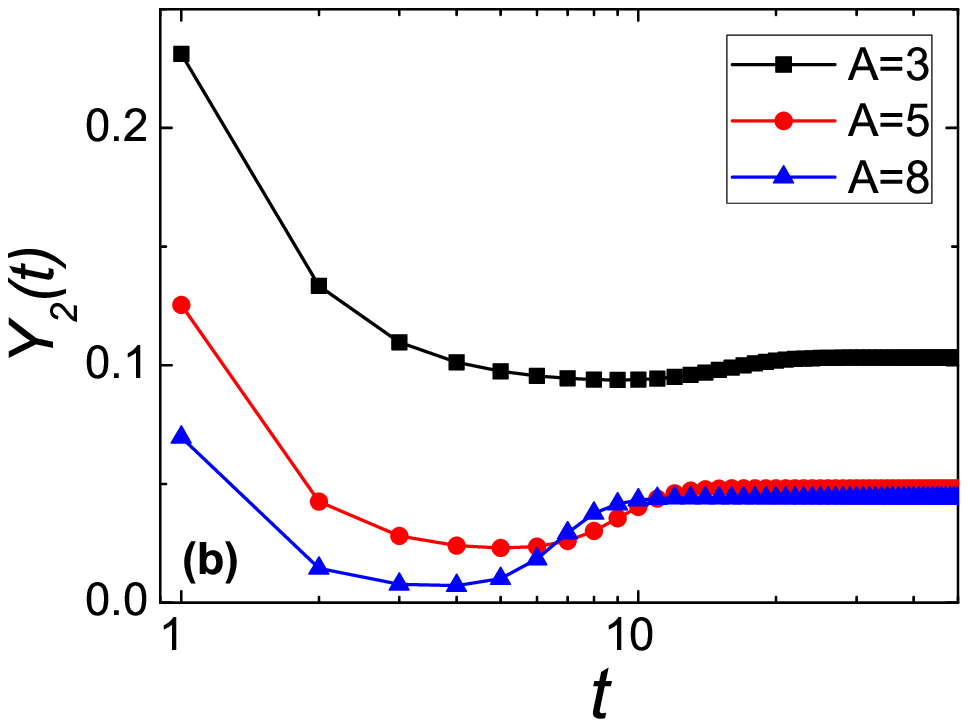}}
\caption{\label{fig:epsart}(color online) Time behavior of the
average degree of the newly recovered nodes (a) and inverse
participation ratio $Y_2$ (b) in BA networks of size $N=5000$.}
\end{figure}

In Fig. 3(a), we plot this quantity for BA networks as a function
of the time $t$ for different values of $A$ and find a
hierarchical dynamics with three plateaus. We can find that all of
the curves show an initial plateau (see also a few previous works
on hierarchical dynamics of the epidemic spreading
\cite{M1,Barthelemy2005,Yan2005} ), which denotes that the
infection takes control of the large degree vertices firstly. Once
the highly connected hubs are reached, the infection pervades
almost the whole network via a hierarchical cascade across smaller
degree classes. Thus, $\langle k_{R}{(t)}\rangle$ decreases to a
temporary plateau, which approximates $\langle k\rangle=2m$. At
last, since the previously infected nodes recovered, all of which
can be regarded as the barriers of spreading, the infection can
only propagate to the smallest degree class. Then, the spreading
process stops fleetly once the infected nodes are all recovered,
as illustrated that $\langle k_{R}{(t)}\rangle$ decreases to zero
rapidly.

Furthermore, we present the inverse participation ratio $Y_2(t)$
\cite{21} to indicate the detailed information on the infection
propagation. First we define the weight of recovered individuals
with degree $k$ by $w_k(t)=R_k(t)/R(t)$. The quantity $Y_2(t)$ is
then defined as:
\begin{equation}
Y_2(t)=\sum_{k}{w_k^2(t)}.
\end{equation}
Clearly, if $Y_2$ is small, the infected individuals are
homogeneously distributed among all degree classes; on the
contrary, if $Y_2$ is relatively larger, then the infection is
localized on some specific degree classes. As shown in Fig. 3(b),
the function $Y_2(t)$ has a maximum at the early time stage, which
implies that the infection is localized on the large degree $k$
classes, as can also inferred from Fig. 3(a). Afterwards $Y_2(t)$
decreases, with the infection progressively invading the lower
degree classes, and providing a more homogeneous diffusion of
infected vertices in the various degree classes. And then,
$Y_2(t)$ increases gradually, which denotes the capillary invasion
of the lowest degree classes. Finally, when $Y_2(t)$ slowly comes
to the steady stage, the whole process ends.

\section{Conclusion}
In this paper, we investigated the behaviors of SIR epidemics with
an identical infectivity $A$. In the standard SIR model, the
capability of the infection totally relies on the node's degree,
and therefore it leaves some practical spreading behaviors alone,
such as in the pear-to-pear, sexual contact, Gmail server system,
and marketing networks. Accordingly, this work is of not only
theoretic interesting, but also practical value. We obtained the
analytical result of the threshold value $\lambda_c=1/A$, which
agree well with the numerical simulation. In addition, even though
the activity of hub nodes are depressed in the present model, the
hierarchical behavior of epidemic spreading is clearly observed,
which is in accordance with some real situations. For example, in
the spreading of HIV in Africa \cite{Bai2006}, the high-risk
population, such as druggers and homosexual men, are always
firstly infected. And then, this disease diffuse to the general
population.

\begin{acknowledgments}
BHWang acknowledges the support of the National Basic Research
Program of China (973 Program) under Grant No. 2006CB705500, the
Special Research Founds for Theoretical Physics Frontier Problems
under Grant No. A0524701, the Specialized Program under the
Presidential Funds of the Chinese Academy of Science, and the
National Natural Science Foundation of China under Grant Nos.
10472116, 10532060, and 10547004. TZhou acknowledges the support
of the National Natural Science Foundation of China under Grant
Nos. 70471033, 70571074, and 70571075.
\end{acknowledgments}


\begin{thebibliography}{ref1}
\bibitem{ex} Note that, if the connections of the underlying networks are localized
(e.g. lattices), then the mean-field approximation is incorrect
and the threshold value is not equal to $1/A$.
\end{thebibliography}
\end{document}